%% file: ms_asp.tex
\documentclass{emulateapj}

\shortauthors{Oda and Totani}
\shorttitle{Deciphering the CSFH from SN survey}

\begin{document}

\title{Deciphering the cosmic star formation history and
the Nature of Type Ia Supernovae by Future Supernova Surveys} 

\author{Takeshi Oda and Tomonori Totani}

\affil{Department of Astronomy, School of Science, Kyoto University,
Sakyo-ku, Kyoto 606-8502, JAPAN}

\email{takeshi@kusastro.kyoto-u.ac.jp}

\begin{abstract}
We investigate the prospects of future supernova searches to get
meaningful constraints about the cosmic star formation history (CSFH)
and the delay time of type Ia supernovae from star formation ($\tau_{\rm
Ia}$), based only on supernova data. Here we parameterize the CSFH by two
parameters, $\alpha$ and $\beta$ that are the evolutionary indices
[$\propto (1+z)^{ \{\alpha , \beta \} }$] at $z \gtrsim 1$ and $\lesssim
1$, respectively, and quantitatively examined how well the three
parameters ($\alpha$, $\beta$, and $\tau_{\rm Ia}$) can be constrained
in ongoing and future supernova surveys. We found that the type
classification of detected supernovae down to the magnitude of $I_{AB} \sim
27$ is essential, to get useful constraint on $\beta$. The parameter
$\tau_{\rm Ia}$ can also be constrained within an accuracy of $\sim$
1--2 Gyr, without knowing $\alpha$ that is somewhat degenerate with
$\tau_{\rm Ia}$.  This might be potentially achieved by ground-based
surveys but depending on the still highly uncertain type-classification
by imaging data. More reliable classification will be achieved by
the \textit{SNAP} mission.
The supernova counts at a magnitude level of $I_{AB}$ 
or $K_{AB} \sim 30$ will allow us
to break degeneracies between $\alpha$ and $\tau_{\rm Ia}$ and
independently constrain all the three parameters, even without knowing
supernova types. This can be achieved by the \textit{SNAP} and \textit{JWST}
missions, having different strength of larger statistics and reach to
higher redshifts, respectively. The dependence of observable quantities on
survey time intervals is also quantitatively calculated and discussed.
\end{abstract}

\section{INTRODUCTION}
It is important to measure the cosmic supernova (SN) rate at present as
well as at high redshift, to understand the cosmic star formation
history (CSFH) and to get a hint for the origin of the type Ia
supernovae (SNe Ia). Although the CSFH up to $z \sim 5$ has been
intensively studied based on star formation rate (SFR) indicators of
high-$z$ galaxies (Lilly et al. 1996; Madau et al. 1996; see Totani \&
Takeuchi 2002 for a recent compilation of observed data by various
methods), the derived CSFH is still a matter of debate since the
estimated SFR densities are considerably different depending on authors
and methods.  The CSFH inferred from SN rate evolution includes
completely independent and complementary information. CSFH estimated by
galactic SFR, especially at higher redshifts, is inevitably biased
toward bright galaxies; in many studies SFR was corrected by
extrapolating the galaxy luminosity function to the highly uncertain
faint end, well below the observational limit.  However, SNe in any
galaxy including dwarfs, or even those in truly intergalactic space, can
equally be detected by SN surveys and hence it is completely free from
this bias. The galactic SFR indicators and core-collapse SN rate
generally trace the massive star formation, but the SN Ia rate gives
some information for low or intermediate mass (typically a few solar
mass) star formation (e.g., Yungelson \& Livio 2000). The delay time of SNe
Ia from star formation to the explosion may be inferred from
observations, giving a useful hint for the still unknown progenitors of
SNe Ia.

There are a number of publications about theoretical predictions of
high-$z$ SN rate as a probe of CSFH and the progenitor of SNe Ia
\citep{Totani96,MaVa98,YuLi98,YuLi00,GiNu99,SuEl00,Kobayashi00}. In
these papers, only the time evolution of the cosmic SN rate density,
which is simply converted from CSFH, was shown.  However, to compare
with observed data it is necessary to predict the expected number of SNe
limited by a given sensitivity of an observation. A work in this
direction was first presented by \citet[hereafter DF99]{DaFr99}.

On the other hand, in recent years there was a remarkable progress about
measurement of type Ia and core-collapse supernova 
rate, pushing out the maximum distance from $z \sim
0$ to $\sim 1.0$
\citep{PaHo96,CaEv99,Hardin00,GaMa02,SDSS03,GaMa04,GOODS04a}.  Some
interesting implications on the CSFH and the Ia delay time have already
been obtained by analyses of these data
\citep{PaFa02,GaMa03,MaGa04,Blanc04,GOODS04b}, though statistics is not large
enough to derive decisive conclusions. For example, one must use
information on CSFH derived by other methods than SNe to derive a strong
constraint on the Ia delay time. Future SN surveys will provide
much better constraints and may allow to determine the interesting
parameters only by supernova information.

However, the cosmic SN rate evolution is dependent on a number of
parameters that are often degenerate with each other, and it is still
unclear how to get useful information from future SN surveys. The
purpose of this paper is to examine what can be learn from future SN
surveys, and find the best strategy of analyzing data to get useful
information. We set our goal as to constrain the three interesting
parameters only based on information obtained by supernova surveys: the rate
of increase of the cosmic SFR from $z=0$ to $\sim 1$, the evolution of
the SFR beyond $z \sim 1$, and the SN Ia delay time, by breaking 
degeneracy between these parameters. We will make predictions on
various observational quantities for future surveys and discuss how well
these parameters can be constrained in the near future.

SNe are detected by variability in more than two epochs of observations.
If a reference frame that is temporally well separated from new
observations is available, we can measure the real SN flux relative to
the reference frame. However, the first SN candidates are often selected
from a flux variability on a shorter time baseline that is comparable to
the expected SN time scale (i.e., month), to find SNe at early
brightening phase, and to avoid contamination of active galactic nuclei
(AGNs) having longer variability time scales. The latest deep SN survey
by GOODS was also based on a short time variability. This selection
process should have some effect on the event rate and hence on SN rate
studies, even though the real flux and light curve can be measured by
using the reference frame in a distant past. Therefore, a prediction of
expected SN counts as a function of variability flux limits for a given
observing time interval, rather than as a function of the real SN flux
in a single-epoch snapshot, would be useful to quantitatively understand
the dependence of the event rate on the variability time scale.  This
was not quantitatively presented in earlier publications, and here we
will present predictions in such a form and discuss possible effects of
choosing different observing intervals.

The plan of this paper is as follows. In \S 2 we describe basic
formulations and calculation methods, and in \S 3 we present the results
of SN detection rate with some discussions on the choice of the
observing intervals. In \S 4 we propose the strategy to decipher the
CSFH and the Ia delay time, and discuss the feasibility of our strategy
referencing to the plans of future proposed SN surveys.  Discussions and
conclusions will be presented in \S 5.  Throughout this paper, the
standard $\Lambda$CDM universe is assumed with following values of the
cosmological parameters: $\Omega_{M} = 0.3$, $\Omega_{\Lambda} = 0.7$,
$H_{0} = 70$ km s$^{-1}$ Mpc$^{-1}$.  The AB magnitude system is used
for the zero-point of photometry, unless otherwise stated.

%*****************************************************
%*****************************************************
%**********************************************section

\section{FORMULATIONS AND CALCULATION METHODS}

\subsection{Basic Equations}  

There are a variety of supernova searches with different search methods
and sampling times, but the selection process of supernovae can
essentially be considered as a search of transient objects with a
sensitivity magnitude limit to flux variability, $m_{\lim}$, between two
epochs separated by a time interval $T$. In reality a more complicated
selection procedures (e.g., more than two epochs) may be taken, but it
can be expressed by a combination of multiple two-epoch selections.  The
total number of detectable SNe, $N(m_{\lim}, T)$, in a survey observing
an area of $\Omega_{\rm obs}$ is calculated from the following equation:
\begin{equation}\label{eq1}
 N(m_{\lim}, T) = \sum_{i} 
\int_{0}^{\infty} \frac{d N_{i}(m_{\lim},T,z)}{dz} dz \ ,
\end{equation} 
where
\begin{equation}\label{eq2}
\frac{ dN_{i}(m_{\lim}, T, z)}{dz} = \Omega_{\rm obs}
\ \frac{r_{i}(z)}{1+z} \  \frac{dV(z)}{dz} 
 \ f_{i}(z,m_{\lim},T) \ .
\end{equation}
Here the subscript $i$ denotes supernova types and the summation in
Eq. (\ref{eq1}) is over all the types of SNe.  The comoving density of
SN rate is described by $r_{i}(z)$, where the rate is measured by the
cosmic time (or the restframe time at $z$). The factor of $(1+z)^{-1}$
accounts for the cosmological time dilation.  The comoving volume
element $dV(z)/dz$ per unit solid angle is in the ordinary definition,
and $f_{i}(z,m_{\lim},T)$ is a function called effective visibility time
(EVT) or control time, which is defined by the observer's time
(at $z=0$). This
function calculates a time duration in which a SN can be detected, and
it depends on the spectra and light curves as well as the magnitude
limit, observing interval, 
and redshift. In a prediction for a snapshot observation without
taking into account variability,
EVT is a duration of a part of SN light curve when a supernova is
brighter than the magnitude limit. On the other hand, in a
variability-limited survey with an interval of $T$, the EVT is the time
duration when the variability flux is brighter than the magnitude limit
for variability, i.e.,
\begin{equation}
f_i = \int  H(m_{\lim} - m_{\rm var}) \ dt \ ,
\end{equation}
where $t$ is a time coordinate for supernova evolution and
$H(x)$ is the step function [$H(x) = 1$ and 0 for $x\geq0$ and $<0$,
respectively], and
\begin{equation}
\label{sub_mag}
m_{\rm var}(z,t,T) = - 2.5 \log \left\{
\left| 10^{-0.4 m(z,t)} - 10^{-0.4 m(z, t+T)} 
\right| \right\}
\end{equation}
is the magnitude of the variability flux between two
epochs of $t$ and $t + T$. Here $m(z, t)$ is the real magnitude of a SN
in a given waveband at redshift $z$ and epoch $t$. It should be noted
that $t$ is the time for observer and the cosmological time dilation
must be included in calculation of $m(z, t)$.

%---------------------------------------------@
\subsection{Light Curves and Spectra of SNe}

\begin{figure*}
\epsscale{1.0}
\plotone{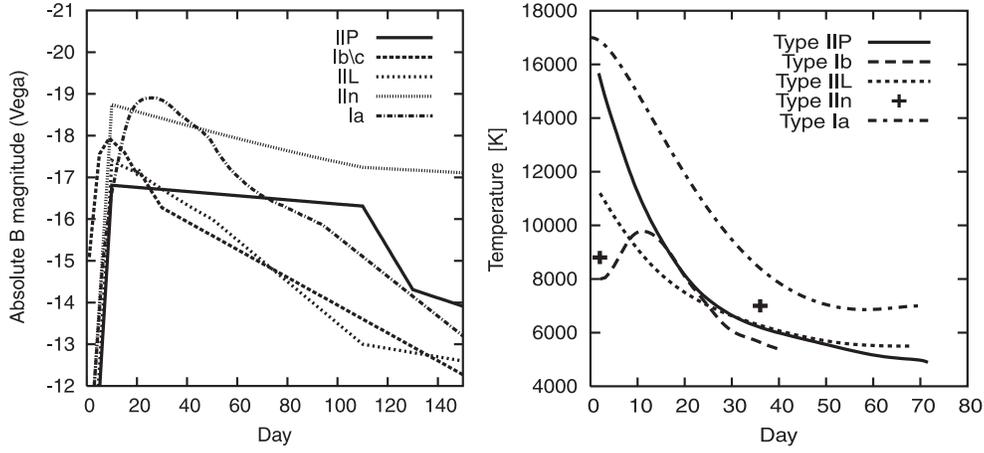}
\caption{Lightcurve and temperature templates for SNe of various types.
The left panel is for the lightcurves used in our calculation and the
right panel is temperature obtained by blackbody-fits to spectra. These
are taken from SN 1999em\citep{HaPi01}, 1979C \citep{BrFa81}
and 1995G \citep{PaTu02} for SN types of IIP, IIL and IIn.
For Ib/Ic,  these are constructed based on the data of 1994I, 1999ex 
and 2002ap \citep{RiDy96,StHa02,Foley03}.
The data for SNe Ia is from \citet{NuKi02}.}
\label{fig:lightcurve}
\end{figure*}

The apparent SN magnitude $m(z, t)$ can be calculated if redshift, light
curve, and spectral evolution are known, taking into account the
K-correction and cosmological time dilation by a standard manner.  We
mostly followed the prescription by DF99 for the calculation of
supernova light curves and spectra, but we collected and used the latest
data published after 1999 if available, to make the most reliable
prediction made so far. Here we only describe these new data, and other
ingredients that are not mentioned here are the same with DF99.
The lightcurves and blackbody temperatures 
used in our calculation are shown in Figure
\ref{fig:lightcurve}.  

The spectra and light curves are taken from \citet{NuKi02} for type Ia
SNe, while those of SNe 1999em \citep{HaPi01}, 1979c \citep{BrFa81}, and
1995G \citep{PaTu02} are used for types IIP , IIL, and IIn,
respectively. The light curves of type Ib and Ic SNe are similar, and we
treat these two as a single population, for which the light curve model
is constructed based on the data of 1994I, 1999ex and 2002ap
\citep{RiDy96,StHa02,Foley03}.  The peak magnitudes differ considerably
from supernova to supernova even within a type, and this is taken into
account by integrating the peak luminosity function of SNe for all
types, assuming a Gaussian form in magnitude with mean magnitudes and
dispersions taken from local values measured by \citet{RiBr02} (shown in
Table \ref{table:abs_mag}).  We use the same relative proportions of the
four subclasses of core-collapse SNe with DF99, which are assumed to be
constant with the cosmic time.  The sum of the relative proportions is
not unity because we do not include peculiar SN1987A-type supernovae,
whose peak magnitude is much fainter ($M_B \sim -15$) and hence they
contribute little to detection number of high-$z$ SNe.  For SNe Ia, the
well-known peak-luminosity and lightcurve-shape relation
\citep{Phi93,Phi99,Perl97, GOODS04b} is taken into account, while the
same shape of lightcurves are used for core-collapse SNe for all peak
magnitudes. The color/lightcurve-shape relation of SNe Ia 
\citep{Phi99} is not included in our calculation. We found that
this effect would hardly change the results presented in our paper.

We do not take account of the Galactic extinction, which can be easily
corrected by the known extinction map. On the other hand, extinction in
host galaxies should be dependent on their various properties and
location of SNe in them, and there should be a large variety 
\citep{Hatano98,Totani99}.  Since it is difficult to
construct a realistic model for this, we simply assume a mean reddening
of $E(B-V) = 0.1$ and the standard Milky-Way type extinction curve
\citep{CaCl89} to make our prediction more realistic than no extinction
case.  (A detailed discussion on this issue will be given in Section
\ref{sec:disc}.)

\subsection{the Cosmic Star Formation History}
\label{section:CSFH}
The CSFH must be specified to predict SN rate as a function of
redshift.  Here we use a phenomenological CSFH model \citep{GaMa04}
inferred from recent
observations of high-$z$ galaxies: 
\begin{equation}\label{eq:CSFH}
\Phi (z) = \frac{2^{0.2} \ \Phi(1.2)}{ \left[ \left(\frac{2.2}{1+z}\right)^
{5\alpha} + \left(\frac{2.2}{1+z}\right)^{5\beta} \right]^{0.2}}  \ .
\end{equation}
Here $\alpha$ and $\beta$ are the SFR evolutionary indices at high 
($z \gtrsim 1$) and low ($z \lesssim 1$) redshift, respectively.  There is
a consensus that the cosmic SFR increases from $z=0$ to $\sim$ 1--2,
while the estimated values differ as $\beta \sim$ 1--3 
\citep[e.g.,][]{LiFe96,SuTr00,Wil02,LaYa02,Galex04}. It is
even more uncertain whether the SFR density is roughly constant to $z
\sim 5$ or declines beyond $z \sim $ 1--2 \citep[e.g.,][]{MaFe96,Hugh98}. 
 In our model, a cut off at high redshift $z_u=10$
is introduced, beyond which there is no star formation, though this cut
off hardly affects our results.

The supernova rate density is calculated by the usual manner as in
earlier papers \citep[][DF99]{MaVa98}. 
The rate of core-collapse SNe
(all types except Ia) is simply proportional to the CSFH, but type Ia
rate is converted from the CSFH by convolving the delay function, i.e.,
the distribution function of the delay time $t_D$ 
from star formation to the type
Ia events\footnote{In some earlier publications, the delay
time $t_D$ is defined as the time from the end of the main-sequence
phase to the occurrence of SNe Ia. However, such definition is
dependent on stellar IMF, and here we define it from the time of
star formation. For a typical IMF
(the Salpeter's  function with the cut-offs at 0.1 and 100 M$_{\odot}$),
the mean difference between the two definitions weighted
over stellar mass is about 0.3 Gyr.}. 
We mainly use a standard exponential form for the delay
function, $f_D(t_D) \propto \exp(-t_D/\tau_{\rm Ia})$
\citep[e.g.][]{MaVa98,GaMa04,GOODS04b} in this paper.  Only in \S
\ref{sec:determin_alpha} we calculate using the Gaussian type function
defined \citet{GOODS04b} to check the difference between them.

In most earlier papers, the normalization of SN rate history is
determined by the number of supernova events from a unit mass of star
formation, but this is highly uncertain. To avoid this uncertainty, in
this paper we always renormalize the calculated SN rate history to the
local SN rate measured by \citet{CaEv99}. This prescription is also
adequate for our purpose, i.e., deciphering CSFH and $\tau_{\rm Ia}$
only based on supernova data.  The renormalization is done separately
for type Ia and core-collapse SNe, whose local rate density is $3.0
\times 10^{-5}$ and $7.5 \times 10^{-5}$ yr$^{-1}$ Mpc$^{-3}$,
respectively \citep{Cap03}.  Therefore the SN detection number
calculated in this paper can be considered as a ratio to the local SN
rates. It should be noted that, with the formulations above, 
the functional form of stellar initial
mass function (IMF) does not affect the SN rate calculation,
provided that the IMF is universal and constant with time.

%------------------------------------------------------@
%------------------------------------------------------@
\section{The Supernova Detection Rate and Observing Time Intervals}

In this section we present some results on the expected SN detection
rate. To see the dependence on the model parameters that we are
interested in, we use two CSFH models and two values of the Ia delay
time, which are largely different but still in the reasonable range.
For one CSFH model, called the ``peak'' model, $(\alpha, \beta) = (-1,
3)$ are used, while $(\alpha, \beta) = (0, 1)$ are used for the other,
called the ``flat'' model. The Ia delay time is set to $\tau_{\rm Ia} =$
0.3 and 3.0 Gyr, which are referred as ``short'' and ``long'' models,
respectively.

\begin{figure*}
\epsscale{1.0}
\plotone{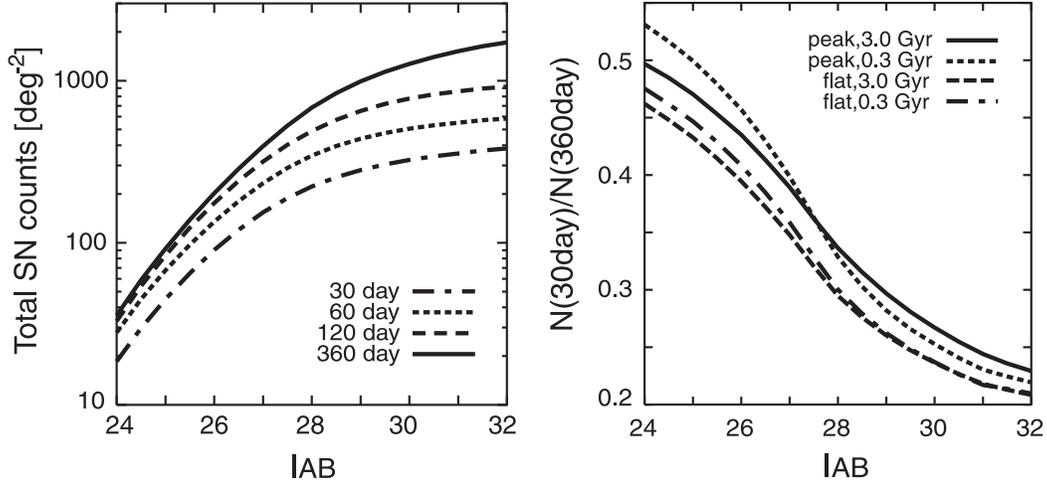}
\caption{ Left~:~Expected number of SNe as a function of limiting magnitude
of variability flux, in a field of 1 deg$^2$. 
Various intervals of observation are assumed as indicated in the figure.
The peak CSFH model and $\tau_{\rm Ia} = 1.0$ Gyr are assumed (see text).
~Right~:~The ratio of expected SN number with an observing interval of
30 days to that of 360 days.
Different line markings correspond
to different models of CSFH (``peak'' or ``flat'') and the Ia delay time
(0.3 or 3 Gyr), as indicated in the panel. 
\label{fig:total_num}}
\end{figure*}

The left panel of Fig.\ref{fig:total_num} shows the total number of
detectable SNe as a function of the magnitude limit for variability
flux, which is a similar plot to popular galaxy counts or star counts as
functions of magnitude limit.  (Therefore we call this plot as ``SN
counts'', hereafter.) The SN counts depend on the interval of
observation; shorter intervals predict less number of detectable SNe.
The long baseline case ($\gtrsim$ 360 days) correspond to the early
studies like DF99, and our results are mostly consistent with those of DF99.

The right panel of figure
\ref{fig:total_num} shows the ratio of SN counts with different
intervals of 30 and 360 days, $N(30)/N(360)$, for different models of
the SN rate history.  The dependence of this ratio on the SN rate
history is not large [$\Delta \{ N(30)/N(360)\} \lesssim 0.03$]. 

\begin{figure*}
\epsscale{1.0}
\plotone{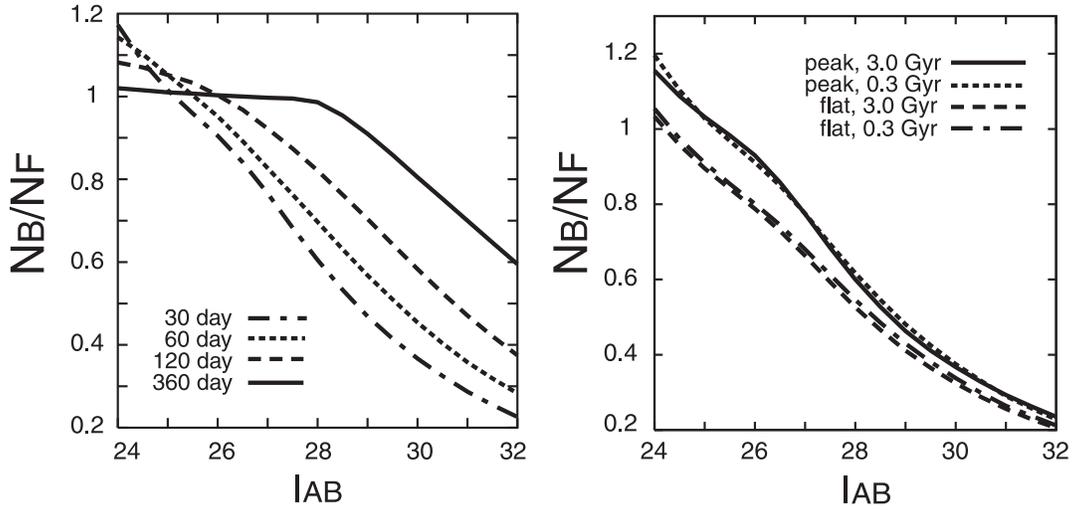}
\caption{Left~:~The number ratio of brightening to fading SNe
between the two epochs of supernova selection. 
Various intervals of observation are assumed as indicated in the figure.
The peak CSFH model and $\tau_{\rm Ia} = 1.0$ Gyr are assumed.
~Right~:~Different line markings correspond
to different models of CSFH (``peak'' or ``flat'') and the Ia delay time
(0.3 or 3 Gyr), as indicated in the panels. The time interval is 30 days.  
\label{fig:Bright_fade}}
\end{figure*}

When SNe are selected by flux variability in two epochs of observations,
SNe can be classified into two categories: SNe whose second-epoch flux
increases or decreases compared with the first epoch. We call these as
brightening and fading SNe. (Therefore it is different from the
derivative of light curves at a given time.)  The ratio of number of
brightening to fading SNe, $N_B/N_F$, is an observable quantity that can
easily be obtained in any supernova surveys, and which is shown in
Figure \ref{fig:Bright_fade}. This ratio is determined by the observing
interval and shape of SN light curves, and potentially it could give
some information of SN type mix.  When the two images are separated
enough, both the number of the brightening and fading phase should be
the same, and the ratio becomes unity. For a fixed interval, the ratio
becomes smaller at deeper magnitude limit, because the faint and long
fading phase of relatively close supernovae mainly contribute to $N_F$.
On the other hand, the ratio becomes larger than the unity at the
brightest limiting magnitude in short interval observations. In this
case only SNe having large variability can be detected, and the initial
rapid brightening phase contributes significantly.  Because of the
combined effects of long decaying phase and cosmological time dilation,
longer intervals are required in deeper observations, to get two
independent reference frames [$N_B/N_F$ = 1].  One year interval is
enough for this in most cases, but it starts to be insufficient for the
deepest surveys possible in the near future ($I_{AB} \gtrsim $ 30).

The right panel of figure \ref{fig:Bright_fade} shows change of
$N_B/N_F$ by different CSFH models and/or type Ia delay time. The
difference is not particularly large and it seems difficult to use the
observational quantities of $N(30)/N(360)$ or $N_B/N_F$ to discriminate
different SN rate histories.  However, these quantities change
significantly with the survey depth and time intervals, which can easily
be measured by observations. Therefore comparing these predictions with
actual data will provide a useful consistency check, such as
examination of possible contamination by non-SN events like AGNs.

%---------------------------------------@
%---------------------------------------@
\section{Deciphering the CSFH and the I\lowercase{a} Delay Time}
\label{section:model-dependence} 

In the framework described above, the three parameters that we want to
constrain by future observations are: (1) the low-redshift CSFH index,
$\beta$, (2) the high-redshift CSFH index, $\alpha$, and (3) the Ia
delay time, $\tau_{\rm Ia}$.  The question that we want to answer is
whether we can constrain these three parameters by future supernova
data, breaking the degeneracy among them.  For this purpose we try the
following strategy.  It is expected that, in relatively shallow surveys,
only supernovae at $z \lesssim 1$ are detectable and the $\alpha$
parameter does not affect significantly the observable quantities,
allowing us to constrain mainly $\beta$ and $\tau_{\rm Ia}$. (The size
of the effect of changing $\alpha$ will quantitatively be checked later.)
The survey time interval is set to 30 days in all the results shown in
this section, and our main conclusions do not depend much on the
choice of the time interval.

\subsection{Determining $\beta$ and $\tau_{\rm Ia}$ by SNe at $z
\lesssim 1$}
\label{sec:determin_alpha}

Then our first approach is to constrain $\beta$ and $\tau_{\rm Ia}$ by
two independent observational quantities at relatively shallow supernova
surveys. The quantity that is the most easily obtained is the total
detected number of SNe including all types at the detection limit of a
survey. For the second observational quantity, we examine the redshift
distribution and the fraction of type Ia supernovae $(f_{\rm
Ia})$. Redshift distribution is easier to obtain by photometric
redshifts of host galaxies without spectroscopy, than $f_{\rm Ia}$ that
requires spectroscopy for a firm determination.  However, we find that
there is a considerable degeneracy between $\beta$ and $\tau_{\rm Ia}$
in the redshift distribution, and the type Ia fraction has a stronger
power to constrain $\beta$ and $\tau_{\rm Ia}$. We use $\alpha = 0$ for
the results presented in this section.

\subsubsection{Redshift Distribution}
\begin{figure*}
\epsscale{1.0}
\plotone{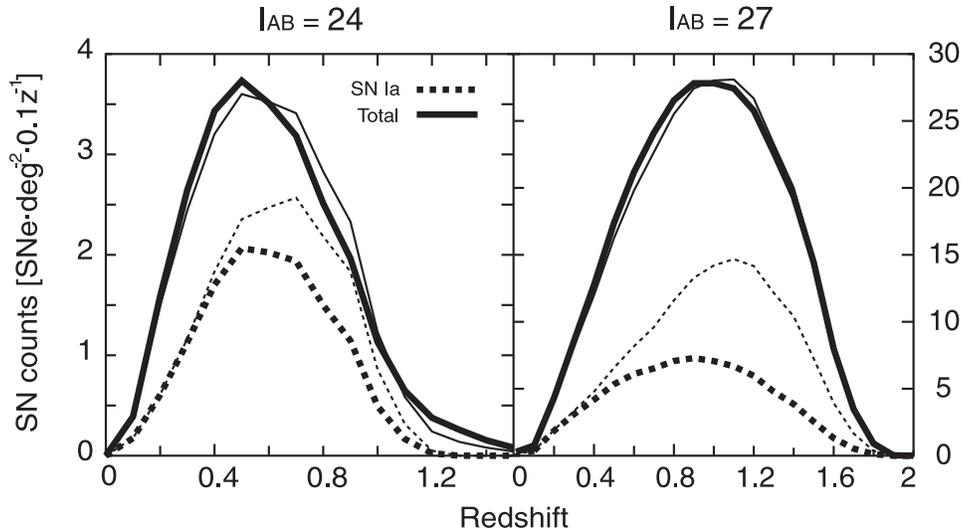}
\caption{Redshift distributions of
SNe in a survey limited by $I_{AB} = 24$ and 27, for two models (thick
and thin curves) whose total expected SN number is almost the same but
the CSFH parameters and type Ia delay time are largely different. The 
parameters are ($\beta$, $\tau_{\rm Ia}$[Gyr]) = (3.0, 4.0) for thick lines 
and (2.3,0.3) for thin lines. The 
solid and dashed lines show the distribution of total number including
all SN types and that of SNe Ia, respectively. 
\label{fig:Reddis_sf}}
\end{figure*}

In order to see the power of breaking degeneracy by redshift
distribution, we calculate the redshift distributions and type Ia
fractions using two cases where the values of $\beta$ and $\tau_{\rm
Ia}$ are very different but the total expected number of SN detection is
the same.  Figure \ref{fig:Reddis_sf} shows the results, where the used
parameters are ($\beta$, $\tau_{\rm Ia}$[Gyr]) = (3.0, 4.0) and (2.3,
0.3).  It can be seen that there is only a small difference in the
redshift distribution of SNe of all types.  The mean redshift differs
only by $\sim$ 0.2, and it is comparable with the accuracy of
photometric redshifts of galaxies \citep[e.g.][]{Fu00}.  Even if
accurate spectroscopic redshifts are available, such a small difference
of distribution may be hidden by uncertainties in theoretical modeling.

Gal-Yam \& Maoz (2004, hereafter GM04) investigated how the three
parameters ($\alpha$, $\beta$, $\tau_{\rm Ia}$) can be constrained by the
redshift distribution of about 1,000 SNe in future surveys, and they
concluded that the redshift distribution gives some useful constraint,
which appears somewhat different from our findings here. In fact, there
is no discrepancy; we made a similar analysis to GM04, and confirmed
that our results are mostly similar to those in GM04. Here we note that
GM04 checked only one parameter set, $(\alpha, \beta, \tau_{\rm Ia}) =
(-2, 4, 1)$, as the input parameters for simulations. 
On the other hand, we found the above degeneracy of the
redshift distribution by using smaller values of $\beta$. This suggests
that the degeneracy becomes more serious than found by GM04 for smaller
values of $\beta$. We also note that, even for the input parameter set
chosen by GM04, the degeneracy between the three parameters cannot be
completely broken, though some parameter space can certainly be ruled out
(see Fig. 7 of GM04). We consider that the redshift distribution only is
not sufficient to measure all the three parameters by breaking the
degeneracy completely.

\subsubsection{The Type Ia Fraction}
\begin{figure*}
\epsscale{1.0} 
\plotone{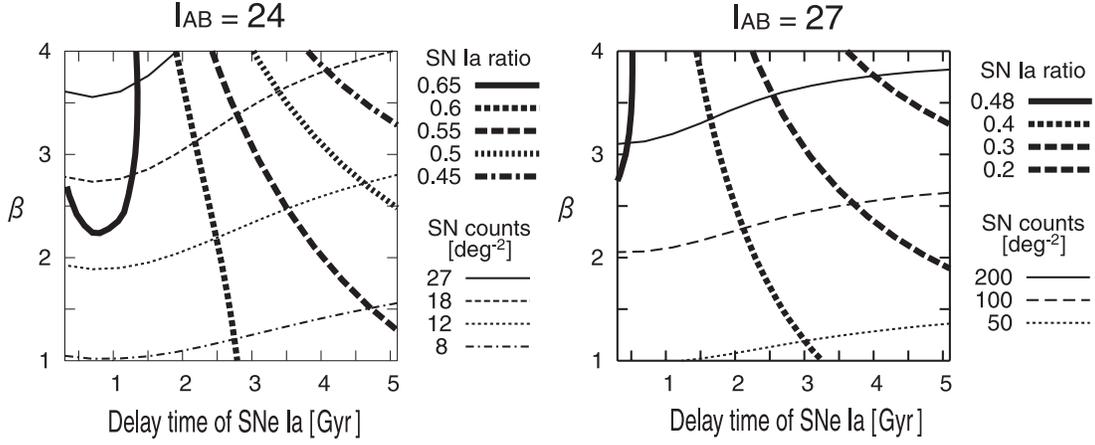} \caption{Contour maps of the
detectable SN number (thin curves) in a square-degree field and the
fraction of SNe Ia (thick curves), for a survey with limiting magnitude
of $I_{AB} = 24$ and 27.  The x-axis is the delay time of SNe Ia
$(\tau_{\rm Ia})$ and the y-axis is the low-redshift evolutionary index
of the CSFH ($\beta$ in Eq. \ref{eq:CSFH}). The search interval
is assumed to be 30 days. If the interval is 360 days, 
the SN counts become about twice the values indicated in the figure,
while the SN Ia fraction changes very little (at most by $\sim$0.05).
\label{fig:cont_totnum} }
\end{figure*}

On the other hand, the difference of type Ia fraction is much larger
than the redshift distribution in Fig. \ref{fig:Reddis_sf}, indicating
that this quantity is a better tool to constrain the parameters of
interest.  For the purpose of seeing this clearer, we made contour maps
of the expected SN detection number and type Ia fraction, as a function
of the two parameters of $\beta$ and $\tau_{\rm Ia}$, which is shown in
Fig \ref{fig:cont_totnum}.  The contours of the detectable SN number and
Ia fraction are mostly perpendicular to each other, demonstrating that
they are good sets of observable quantities to break the degeneracy.
Currently $I_{AB}$ magnitude of $\sim $ 24 is the spectroscopic limit
for the largest ground-based telescopes, and it is difficult to measure
the SN types beyond this magnitude.  In this case the expected
difference of Ia fraction is not very large ($f_{\rm Ia} \sim$ 0.5--0.6
for an interesting range of $\tau_{\rm Ia}$).  On the other hand, future
space borne spectroscopy or extremely large ground-based telescopes (30
m class) would allow the type classification at much deeper magnitudes. If
SN types at $I_{AB} \sim 27$ are reliably determined, we expect that the
Ia fraction differs by a factor of about 2 for the same range of
$\tau_{\rm Ia}$.

In this analysis we have assumed that the effect of $\alpha$ on our
predictions for shallow surveys is small. If the SN Ia fraction is
reliably measured, the parameter $\beta$ can be estimated without
knowing $\alpha$, since the core-collapse SN rate evolution to $z \sim 1$
simply reflects the SFR evolution. However, uncertainty of $\alpha$
affects the estimate of $\tau_{\rm Ia}$.  We found that, when the
$\alpha$ is varied from $-2$ to 1, the estimate of $\tau_{\rm Ia}$ by the
same observational quantities is changed by 1--2 Gyr for the limiting
magnitude of $I_{AB} =$ 24--27. This degeneracy can be broken by using
deeper SN counts at magnitude of $\gtrsim 30$ (see below).

\subsubsection{On the Form of the Ia Delay Function}
The exact form of delay function is unknown, though there are some
theoretical models \citep{YuLi98,YuLi00}. \citet{GOODS04b} found that
the Gaussian form of the delay function, $f_D(t) \propto
\exp[-(t-\tau_{\rm Ia})2/ (2\sigma_{\rm Ia}^2)]$ with $\sigma_{\rm Ia}
\sim (0.2$--0.5)$\tau_{\rm Ia}$, fits better than the exponential form
used so far in this paper.  Therefore we repeated the above calculations
but changing the delay function into the Gaussian form, to estimate the
error on $\tau_{\rm Ia}$. It changes by about 1 Gyr, for the limiting
magnitude of $I_{AB} \sim 27$.  However, as shown by \citet{GOODS04b},
the difference of the delay function appears most clearly in the
redshift distribution of SNe, and hence the degeneracy with the delay
function form may be removed by using the redshift distribution.

\subsection{Determining $\alpha$ by SNe beyond $z \gtrsim 1$}

It is obvious that we have to observe SNe well beyond $z \sim 1$ to
determine $\alpha$, and detection limit of $I_{AB} \sim 27$ is still not
sufficient for this purpose (see Fig. \ref{fig:Reddis_sf}).  Future
projects such as \textit{SNAP} and \textit{JWST} can detect supernovae down to
$\sim 30$ magnitudes, and their data will be useful to constrain
$\alpha$, though it may be difficult to determine the SN types at such
faintest magnitude. 

\begin{figure*}
\epsscale{1.0}
\plotone{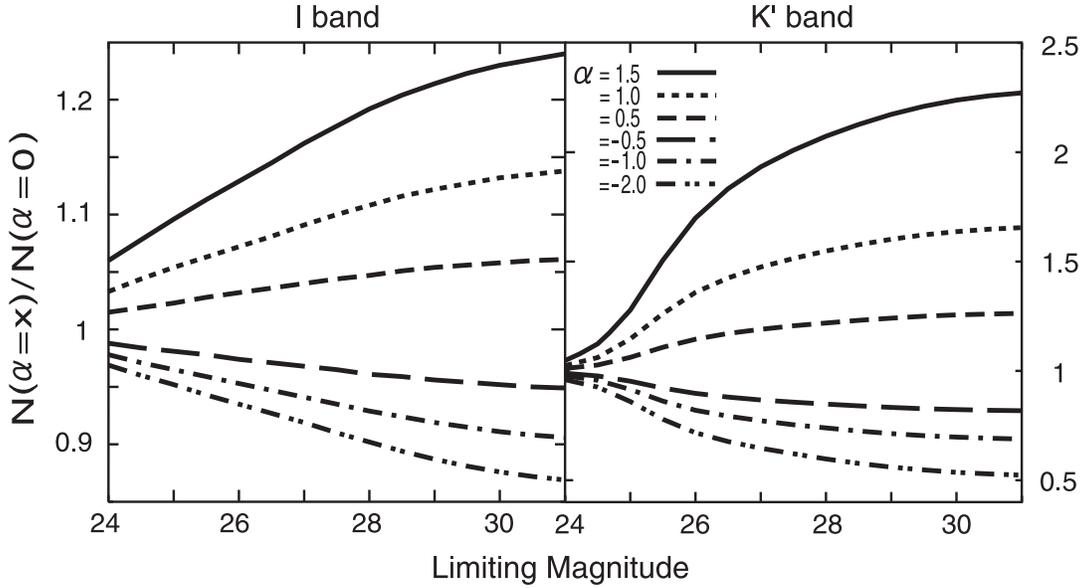}
\caption{Detectable SN number normalized by the case of $\alpha = 0$ 
as a function of limiting magnitude, for all SN types.  
Various $\alpha$ are assumed as indicated in the figure. 
The used parameters are ($\beta$, $\tau_{\rm Ia}$[Gyr]) = (2.0, 1.0).
Left and right panels are for I and K' band, respectively. 
\label{fig:sfh_beta}
}
\end{figure*}

We calculate the expected SN counts varying the $\alpha$ parameter and
the results are shown in Fig. \ref{fig:sfh_beta}.  In this calculation
we choose a typical parameter set of ($\beta$, $\tau_{\rm Ia}$) = (2.0,
1.0).  As expected, the change of SN counts by different $\alpha$ becomes
larger for deeper observation. It should also be noted that the
difference is much larger in the $K$ band than $I$ band, since the
redshift distribution extends farther in redder bands ($z \sim$ 2 for
$I$ and $z \gtrsim 5$ for $K$ at $\sim 30$ mag).  In a $K$-band
observation with $m_{\rm lim} = 30$, the fraction of SNe beyond $z =2$ is
30--40 \% in the case of $\alpha = 0$. 

This is a calculation for fixed values of $(\beta, \tau_{\rm Ia})$.  By
a supernova survey down to $I_{AB} \sim 27$ with secure type
classification, the parameter $\beta$ can be constrained well by using
core-collapse SN statistics, while the degeneracy between $\tau_{\rm
Ia}$ and $\alpha$ remains. It is then possible to break the degeneracy by
finding a set of $(\alpha, \tau_{\rm Ia})$ satisfying both the type Ia
statistics to $I_{AB} \sim 27$ and the all-type SN statistics to $\sim $
30 magnitudes.

\subsection{Prospects for the Planned Projects}

The current or past supernova surveys from ground, such as SCP
\citep{PaFa02} and HZT \citep{Tonry03}, have reached a depth of about
$I_{AB} \sim 24$.  The recent GOODS project has achieved a depth of
F850LP $\sim 26$ thanks to HST observations, but high signal-to-noise
spectroscopic observation is still limited to $I_{AB} \sim 24$ by the
largest ground telescopes. As shown above, the type Ia fraction
needs to be determined with an accuracy distinguishing $\Delta f_{\rm
Ia} \sim 0.1$ to determine $\beta$ and $\tau_{\rm Ia}$ by surveys
limited to $I_{AB} \sim 24$.  Requiring statistical signal-to-noise of 5, a
required number of SNe Ia becomes $N \gtrsim 1,000$, which is much
larger than the number obtained so far.  However, the ongoing
ground-based projects such as SNLS \citep{SNLS04} or ESSENCE 
\citep{ESSENCE02,ESSENCE04} will greatly increase the
number of supernova to this magnitude. These surveys could give some
useful constraint on $\beta$ and $\tau_{\rm Ia}$, depending on the
accuracy of $f_{\rm Ia}$ and systematic model uncertainties.

The situation would be improved if we can measure the Ia fraction down to
$I_{AB} \sim 27$, as shown in Fig. \ref{fig:cont_totnum}.  Spectroscopic
type classification to this depth is impossible by current ground-based
facilities, but classification by information obtained by imaging
observations, such as light-curve shapes, colors, or host galaxy
properties, may still be possible, as tried by \citet{GOODS04b} for
the GOODS data. (See also Poznanski et al. 2002.)  SNe II have very
distinctive light-curves (type II-P or II-L), and are typically bluer
than SNe Ia at early times.  However, the reliability of
this kind of approach is still rather uncertain so far; only about two
thirds of the 42 GOODS SNe are secure type identifications
\citep{GOODS04a}.  A major breakthrough in more distant future will be
brought by spaceborne spectroscopy, such as \textit{SNAP}, reliably
measuring $f_{\rm Ia}$ to $I_{AB} \sim 27$.  The \textit{SNAP} will
detect and get high signal-to-noise spectra of a few thousands of type
Ia supernovae to $z \sim$ 1.7 \citep{SNAP04}, which corresponds
to the peak magnitude of $I_{AB} \sim 27$ (see
Fig. \ref{fig:Reddis_sf}). Planned extremely large telescopes 
($\sim$ 30 m class) will also be useful for type classification of the
faintest SNe.

The spaceborne projects such as \textit{SNAP} and \textit{JWST} 
\citep{Panagia03} will also
allow us to break the degeneracy between $\alpha$ and $\tau_{\rm Ia}$
by SN counts at magnitude of $\sim 30$. As shown above, \textit{SNAP}
will have a sufficient number of supernovae but does not probe very high
redshift beyond $z \sim 1$ because of the wavelength coverage. The 
\textit{JWST} could probe higher redshift, but small field of view (14
square arcminutes) does not allow to detect many supernovae at one
snapshot. According to our estimate with the parameter $\alpha = 0.0$ ,
$\beta = 3.0$ and $\tau = 1.0$, \textit{JWST} will find about 
10 SNe down to $K \sim 30$ in one snapshot.  To measure the
supernova counts with an accuracy of $\sim 10$--20\% to obtain meaningful
constraint on $\alpha$, about 25--100 supernovae are required.

%*****************************************************
%*****************************************************
%**********************************************section
\section{Discussion and Conclusion}\label{sec:disc}

In this paper we examined the dependence of observational quantities
obtained by supernova surveys on the CSFH and the delay time of SNe Ia 
($\tau_{\rm Ia}$) in
detail, and tried to identify the best strategy to get meaningful 
constraints on them by planned future projects. Specifically, we chose
three parameters of $\alpha$, $\beta$, and $\tau_{\rm Ia}$, and
investigated how well these parameters can be constrained in the future
based only on supernova survey data, 
where $\alpha$ and $\beta$ are the indices of the cosmic SFR evolution
[$\propto (1+z)^{ \{\alpha, \ \beta \}}$] at $z \gtrsim$ 1 or $\lesssim$
1, respectively.

Relatively shallow surveys ($I_{AB} \lesssim 27$) are relevant mainly
with the parameters of $\beta$ and $\tau_{\rm Ia}$.  We have shown that
it is essential to reliably measure the number fraction of SNe Ia
($f_{\rm Ia}$) in a survey sensitive to $I_{AB} \sim 27$, in order to
measure these parameters.  If a $\sim 20$\% difference in $f_{\rm Ia}$
is clearly distinguished for supernovae at magnitude of $I_{AB} \sim
27$, a useful constraints on $\beta$ will be obtained by core-collapse
SN rate evolution.  The constraint on $\tau_{\rm Ia}$ is somewhat
degenerate with the unknown $\alpha$, but still it can be constrained
within an accuracy of $\sim$ 1--2 Gyr. Some regions of the parameter
space of $\beta$ and $\tau_{\rm Ia}$ are seriously degenerated for the
redshift distribution, and hence we consider that the Ia fraction is a
better quantity for this purpose. However, the redshift distribution
would be useful to constrain the functional form of the delay function
(e.g., exponential or Gaussian).  

Type classification down to $I_{AB} \sim 27$ may be potentially achieved
by ongoing or near future ground-based surveys such as SNLS and ESSENCE,
but type classification inevitably depends on less reliable methods
based on imaging information.  A more reliable study will become
possible by spaceborne spectroscopy, as proposed for \textit{SNAP},
or future extremely large telescopes on the ground.  The
parameter $\alpha$ can be constrained with breaking the degeneracy with
$\tau_{\rm Ia}$, by ultra-deep SN surveys down to magnitudes of $\sim$
30 without SN type classification, which would become possible by future
projects such as \textit{SNAP} and \textit{JWST}. The former has an
advantage of larger statistics while the latter has an advantage of
reaching higher redshifts by observing in longer wavelength. It should
also be noted that, in the above discussion, we assumed that the local
supernova rates, both for type Ia and core-collapse SNe, are well
measured to give a good normalization that can be directly compared with
high redshift observations. Therefore the effort to improve the accuracy
of the local supernova rate is also highly encouraged
\citep[e.g.][]{NSNF02}.

It should be noted that there is a number of simplifications in the
calculation presented here, e.g., a constant IMF, no evolution in
supernova light curves or spectra, a universal probability of supernova
event occurrence per unit mass of star formation, and no evolution in
the Ia delay time. These uncertainties might induce some bias in the
CSFH or Ia delay time inferred from the SN rate history. On the other
hand, it is possible to get information for these uncertain factors, if
we use the CSFH determined by independent other methods (such as SFR
estimates of high-$z$ galaxies) as an external constraint. 

Dust extinction has been incorporated in the same way for the SNe Ia and
core-collapse SNe, with a fixed amount of extinction and no evolution
with $z$. We adopted this simple treatment because it is quite difficult
to construct a realistic model of dust extinction based on the present
knowledge about galaxy evolution.  However, it is likely that extinction
has a wide distribution even within a galaxy (Hatano, Branch, \& Deaton
1998), and mean extinction should also evolve with galaxy evolution
(Totani \& Kobayashi 1999).  Because the time delay from star formation
to supernova occurrence is different, core-collapse SNe could be much
more heavily extincted by dust than SNe Ia \citep{Mann03}. If the cosmic
star formation is heavily hidden by dust at $z \gtrsim 2$ as inferred
from submillimeter observations (e.g., Hughes et al. 1998), it might be
appropriate to use different values of $\alpha$ for type Ia and
core-collapse SNe. These issues should be kept in mind when the strategy
discussed in this paper is actually applied to future data.  To see a
typical effect of changing extinction, we repeated calculations with no
extinction for SNe Ia and twice larger extinction for core-collapse SNe.
The Ia delay time and $\beta$ are then affected by $\sim$1 Gyr and
$\sim$0.5, respectively.

We presented our calculation of expected supernova counts as a function
of variability flux in a given survey time interval, instead of the real
flux. We also examined quantitatively the change of the detectable
number of supernovae with time interval, and calculated the ratio of
brightening and fading supernovae between the two epochs of the SN
selection, which were not quantitatively reported in the earlier
literature.  These are useful for planning the observing intervals for
the future observations.  These quantities can easily be measured, and
we examined a possibility of using these observational information to
constrain the interesting parameters of CSFH and $\tau_{\rm Ia}$, but
unfortunately they are not particularly useful. However, comparing these
quantities to the theoretical predictions would be an important
consistency check to examine, e.g., a possible contamination of other
objects such as AGNs.

We would like to thank an anonymous referee for many constructive comments
that greatly improved this paper. 
This work has been supported by the Grant-in-Aid for the
21st Century COE "Center for Diversity and Universality in Physics"
and for TT (16740109) from the Ministry of Education, Culture, Sports,
Science and Technology (MEXT) of Japan.

%*****************************************************
%*****************************************************
%***********************************************figure

%\clearpage
 
%\clearpage

\input{table1.tex}

\end{document}

%% file: table1.tex
\begin{deluxetable}{cccc}
\tablecolumns{4}
\tablewidth{0pt}
\tablecaption{ Peak absolute magnitudes (Vega),  their dispersion,
and Relative Proportions of Supernova Types
\label{table:abs_mag}}
\tablehead{
\colhead{Type} & \colhead{M$_{B}$}   & \colhead{$\sigma$} & \colhead{f}}
\startdata
Ia   & -19.13 & 0.56\tablenotemark{a} &  ---\\
Ib/c & -17.71 & 1.39 &  0.23\\
IIP  & -16.67 & 1.12 &  0.30\\
IIL  & -17.70 & 0.90 &  0.30\\
IIn  & -18.82 & 0.92 &  0.02
\enddata
\tablenotetext{a}{The dispersion of SNe Ia is uncorrected about
the light-curve-shape/luminosity relation.}
\tablerefs{\citet{RiBr02} for $M_B$ and $\sigma$, and \citet{DaFr99}
for $f$.}
\end{deluxetable}